\documentclass{jaa}
\usepackage{natbib,hyperref}
\usepackage[normalem]{ulem}
\usepackage{graphicx,color}

\begin{document}\sloppy

%%paper title
%%For line breaks \\ can be used within title
\title{Gravitational physics in the context of Indian astronomy: A vision document}

%%author names are separated by comma (,)
%%use \and before the last author name
%%use a * along with the number separated by comma
%% for the  author for correspondence
%%\textsuperscript{number} is used for affiliation
%%\affilOne, \affilTwo etc., upto \affilTwentyfive is possible
%%Please note the first letter after \affil is capitalised in the command
%%

\author{P. Ajith~\textsuperscript{1}, K. G. Arun~\textsuperscript{2}, Sukanta Bose~\textsuperscript{3}, Sumanta Chakraborty~\textsuperscript{4}, Shantanu Desai~\textsuperscript{5},  A. Gopakumar~\textsuperscript{6}, Sanved Kolekar~\textsuperscript{7}, Rajesh Nayak~\textsuperscript{8}, Archana Pai~\textsuperscript{9}, Sudipta Sarkar~\textsuperscript{10}, Jasjeet Singh Bagla~\textsuperscript{11,17}, Patrick Das Gupta~\textsuperscript{12}, Rahul Kashyap~\textsuperscript{13}, Prashant Kocherlakota~\textsuperscript{14, 15}, Prayush Kumar~\textsuperscript{1}, Banibrata Mukhopadhyay~\textsuperscript{16}}

\affilOne{\textsuperscript{1} International Centre for Theoretical Sciences, Tata Institute of Fundamental Research, Bangalore 560089, India.\\}

\affilTwo{\textsuperscript{2}Chennai Mathematical Institute, Siruseri, 603103, Tamilnadu, India.\\}

\affilThree{\textsuperscript{3}Inter-University Centre for Astronomy and Astrophysics, Post Bag 4, Ganeshkhind, Pune 411 007, India\\}

\affilFour{\textsuperscript{4}School of Physical Sciences, Indian Association for the Cultivation of Science, Kolkata 700032, India.\\}

\affilFive{\textsuperscript{5}Department of Physics, IIT Hyderabad, Kandi, Telangana 502284, India.\\}

\affilSix{\textsuperscript{6}Department of Astronomy and Astrophysics, Tata Institute of Fundamental Research, Mumbai 400005, India\\}

\affilSeven{\textsuperscript{7}Indian Institute of Astrophysics, Bangalore, India.\\}

\affilEight{\textsuperscript{8}Indian Institute of Science Education And Research Kolkata, Mohanpur, Nadia - 741 246 West Bengal, India\\}

\affilNine{\textsuperscript{9}Department of Physics, IIT Bombay, Powai, Mumbai 400076, India.\\}

\affilTen{\textsuperscript{10}Indian Institute of Technology, Gandhinagar, Gujarat 382355.\\}

\affilEleven{\textsuperscript{11}Indian Institute of Science Education and Research (Mohali), Department of Physical Sciences, Sector 81, Sahibzada Ajit Singh Nagar, Punjab 140306, India\\}
\affilTwelve{\textsuperscript{12}Department of Physics and Astrophysics, University of Delhi, Delhi 110007\\}
\affilThirteen{\textsuperscript{13}Institute for Gravitation and the Cosmos and Physics Department, Penn State University, University Park PA 16802, USA\\}
\affilFourteen{\textsuperscript{14}Black Hole Initiative at Harvard University, 20 Garden St., Cambridge, MA 02138, USA\\}

\affilFifteen{\textsuperscript{15}Center for Astrophysics, Harvard \& Smithsonian, 60 Garden St., Cambridge, MA 02138, USA\\}

\affilSixteen{\textsuperscript{16}Astronomy \& Astrophysics Group, Department of Physics, Indian Institute of Science, Bangalore 560012\\}

\affilSeventeen{\textsuperscript{17}National Centre for Radio Astrophysics, Tata Institute of Fundamental Research, Ganeshkhind, Pune 411007, India\\}

%%escape two column mode for title, affiliation and abstract
%%by giving \twocolumn command as shown

\twocolumn[{

\maketitle

%%include \corres to print the corresponding author Email id
%\corres{abc@xyz.com}

%%include \msinfo for
%%manuscript information such as
%%received, revised and accepted dates
%%
%\msinfo{1 January 2015}{1 January 2015}

%%abstract
\begin{abstract}
 Contributions from the Indian gravity community have played a significant role in shaping several branches of astronomy and astrophysics. This document reviews some of the most important contributions and presents a vision for gravity research in the context of astronomy \& astrophysics in India. This is an expanded version of one of the chapters in the recently released Vision Document of the Astronomical Society of India.
 \end{abstract}

%%insert keywords separated by 3 hyphens using \keywords{words}

}]
%%close the twocolumn escape here

%%include \doinum{number}for the DOI number in the header
%%include \volnum{number} for the volume number in the header
%%include \year{yyyy} for  year of publication in the header
%%include \pgrange{num--num} page range of article in the header
%%include \artcitid{num} for the article citation id
%%include \lp to print last page of the article
%%include \setcounter{page}{pagenum} for the exact starting page of the article

\doinum{12.3456/s78910-011-012-3}
\artcitid{\#\#\#\#}
\volnum{000}
\year{0000}
\pgrange{1--}
\setcounter{page}{1}
\lp{1}

\section{Introduction}
The field of gravitational physics has undergone revolutionary changes over the past decade and that has led to three Nobel Prizes~\cite{GWnobel,Cosmonobel,BHnobel} over the past five years. Gravitational Physics is now widely recognized as a branch of observational astronomy that has significantly influenced almost every branch of astronomy. The detection of gravitational waves (GWs) by the LIGO and Virgo observatories~\citep{LIGOScientific:2016aoc}, evidence for nHz gravitational waves from Pulsar Timing Arrays (including Indian Pulsar Timing Array Consortium)~\citep{NANOGrav:2023det,InPTAEPTA}, the high-resolution radio imaging of the shadow regions of supermassive black holes (SMBHs) by the Event Horizon Telescope (EHT; \cite{EHTC+2019a}), and probes of the supermassive BH at the center of our galaxy using precise tracking of stellar orbits~\citep{Ghez:2008ms, Gillessen:2008qv} are some of the path-breaking discoveries of the past decade. 

All of these observational frontiers are expected to improve their science capabilities in the next decade by increasing sensitivity and coverage, providing new powerful probes of fundamental physics, astrophysics and cosmology. These observational quests are closely connected to theoretical research in gravity and related fields, such as magneto-hydrodynamics, high-energy physics, nuclear physics, etc. Thus, future research in gravity requires synergy between large experimental and computational facilities as well as individual-driven theoretical research. This document aims to present a vision for gravity research in the next two decades in the context of Indian astronomy.

\section{Global Status}
The first detection of GWs by LIGO~\cite{LIGOScientific:2016aoc} in 2015 opened up a new branch of astronomy. GW observations provided a new tool~\citep{GW150914TGR} to probe the genuinely strong-field dynamics associated with mergers of black holes (BHs) and neutron stars (NSs), and hence enabled powerful tests of general relativity (GR)~\citep{GW150914TGR}. They uncovered a new population of heavy BHs and provided a new means to characterize binary BH populations in the universe. Another ground-breaking discovery from this field is the multi-messenger observation of a  binary NS merger~\citep{GW170817} GW170817 and an associated gamma-ray burst GRB170817A~\citep{MMA}, followed by a kilonova~\cite{SwopeEMCounterpart}, which was observed in almost every band of the electromagnetic spectrum. This gave us key insights into the central engine of gamma-ray bursts and the structure of their relativistic jets. The first measurement of the Hubble constant independent of the cosmic distance ladder, hints of the origin of heavy elements, a constraint on the dense nuclear equation of state and tests of modified gravity theories in the infrared  are among the most prominent scientific outcomes of this discovery.  

LIGO and Virgo detectors continue to improve their sensitivities and recently the Japanese KAGRA detector has joined the global GW detector network. LIGO-India, a project to build a third LIGO detector in India~\citep{LIGOIndiaProposal:2011,Saleem:2021iwi}, is in progress. There are planned upgrades of the ground-based detectors such as A$^\#$ and Voyager~\citep{LVKLRR}, and proposed third-generation detectors such as Cosmic Explorer~\citep{CE:2019iox} and Einstein Telescope~\citep{ET} that will elevate GW detectors to a novel and precise astrophysical tool in the next 15 years. 

Complementing the ground-based detectors, the International Pulsar Timing Array (IPTA) consortium is expected to inaugurate the nano Hz (nHz) GW astronomy era in the coming years as the current data sets are showing the first evidence for nHz GWs~\citep{NANOGrav:2023det}. The Indian Pulsar Timing Array (InPTA), a constituent of  IPTA,  employs the upgraded Giant Meter Wave Radio Telescope (uGMRT; ~\cite{uGMRT} for contributing to various nHz GW astronomy-related efforts. The IPTA  consortium is expected to detect first a stochastic nHz GW background from a population o merging supermassive BHs that weigh billions of solar masses and later from individual sources with high statistical significance.  Complementing the earth and galaxy-based GW observatories will be the Laser Interferometer Space Antenna (LISA), a mHz GW detector in space that is expected to be launched by the middle of the next decade. Among other things, LISA would be sensitive to mergers of supermassive BHs with masses of millions of solar masses. There are interesting proposals to build deci-Hz GW detectors in space or on the moon that will bridge the gap between mHz LISA and ground-based detectors that operate in the audio frequencies ($\sim 10-10^3$ Hz). Missions that will search for imprints of very low-frequency GWs in the polarisation of the cosmic microwave background (CMB) are also being explored actively. In summary, the next decades is slated to witness the extension of GW astronomy to multiple bands, apart from deeply probing the high-frequency window. These will have major impacts on our understanding of the cosmos in several ways.

Similarly, observation of the supermassive compact object at the center of our galaxy has been a challenge that has been carried over the past decades finally giving us conclusive evidence for its presence predominantly via infrared telescopes. High-resolution radio imaging of the central compact objects in our galaxy as well as M87 by Event Horizon Telescope Collaboration provided us with the first horizon-scale image of the compact object and its accretion disc \citep{EHTC+2019a, EHTC+2022a}. Though these supermassive compact objects are consistent with being BHs (BHs) in GR \citep{EHTC+2019f, EHTC+2022f}, more detailed observations should be required to confirm this with a high level of precision \citep{Psaltis+2020, Kocherlakota+2021, EHTC+2022f, Vagnozzi+2023, Ayzenberg+2023}.  
Next-generation ground-based very long baseline interferometry (VLBI) facilities such as the next-generation EHT (ngEHT; \citealt{Doeleman+2023}) aim to improve UV coverage, flux-sensitivity, dynamic range, number of observing radio bands, and to introduce a new long-term monitoring mode. These upgrades will enable capturing movies of the horizon-scale accretion phenomena and also of the base of the large-scale jet in M87 \citep{Johnson+2023}. This would directly enable probing the physics of flaring events that are thought to be caused by magnetic reconnection in the accretion flow \citep{Ripperda+2022} as well as to test the \cite{Blandford+1977} mechanism that is believed to generate powerful relativistic outflows in active galactic nuclei. Since the latter is an electromagnetic version of the Penrose process \citep{Lasota+2014}, it would provide direct evidence for spin energy extraction from the ergoregions of astrophysical BHs. Observations of electromagnetic counterparts of nHz-GW emitting massive BH binary candidates like OJ 287 will also become possible. Furthermore, space-based VLBI extensions to ground arrays will provide access to extremely high-resolution images and movies of M87* and Sgr A*, and thus enable us to resolve the photon rings of supermassive BHs, which comprise the higher-order images of the horizon-scale accretion flow, caused due to extreme gravitational lensing \citep{Johnson+2020}. Measurement of the diameter and ellipticity of the first-order image would provide a $\lesssim10\%$ measurement of the mass and spin of M87* and Sgr A*.

Inferring the equation of state of NSs is among the most interesting topics in gravitational physics which has implications for astrophysics and nuclear physics. Traditional probes using radio~\citep{Lattimer:2000nx} and X-rays~\citep{Steiner:2012xt} have played a vital role in our current understanding of the topic. The discovery of heavy NSs from radio observations of binary pulsars and the measurement of NS radius from x-ray observations have provided very useful constraints on the nuclear equation of state. GW observations also allow us to measure directly the elastic property of NSs~\citep{Agathos:2015uaa,Foucart:2020ats}, hence, providing us with a new window into the same physical system. More recently, transients, such as kilonovae, associated with NS mergers have proven to be a handy tool to probe the dense state of NSs~\citep{Margalit:2019ssg}. Such joint multimessenger inference of nuclear matter requires precise modelling of gravitational waves, high-energy and electromagnetic transients from the same system around the same time when undergoing the merger process in the source's rest frame. A synergy between these methods is likely to play a crucial role in solving fundamental questions in this field in the future.

Theoretical development on the gravitational physics front, analytical and numerical, has a very crucial role to play in terms of achieving all the above-mentioned goals. The fact that the Physics Nobel Prize for 2020 was shared by R. Penrose, who made pioneering contributions towards the understanding of BHs and their formation, underscores the role theoretical physics can play in major astrophysical discoveries. GW observations need accurate theoretical models of the expected signals, VLBI observations require models of shadows of compact objects, and models of stellar orbits around the galactic center have been used to constrain deviations from GR. There is a vast theoretical physics community, which is  involved in the modeling of BH dynamics within GR and in modified theories of gravity. The results should be of immense importance in achieving the above-mentioned scientific objectives.

\section{Open Questions}
\begin{enumerate}
    \item \emph{Discovering and understanding new GW sources and phenomena:} Upcoming GW observations have the potential to discover spinning NSs, stochastic GW background, core-collapse supernovae, SMBH binaries, extreme mass ratio inspirals, gravitationally lensed GW signals, etc. Some of these can be observed as multi-messenger sources. These observations can also help us to understand the formation, evolution and demography of stellar-, intermediate-mass- and supermassive BH binaries, structure of ultra-relativistic jets, etc.

\item \emph{Probing extreme gravity and extreme matter:} This includes testing GR in the relativistic strong field regime, identifying BH horizons and inferring the state of extremely dense matter and structure of NSs.

\item \emph{Understanding the nature of dark matter and dark energy:} Through CMB, Type Ia supernovae, galaxy surveys, gravitational lensing and GW observations, and complimenting them with theoretical studies including alternative theories of gravity. 
 
\item \emph{Discovering and constraining exotic physics:} Existence of cosmic strings, domain walls, primordial black holes, exotic compact objects and extra space-time dimensions through a variety of astronomical observations. 

\end{enumerate}

\section{Research and status of the field in India}
Starting from the seminal work of N. R. Sen, B. Datt, V. V. Narlikar, P. C. Vaidya, A. K. Raychaudhuri and C. V. Vishveshwara,  Indian scientific community has made substantial contributions to theoretical research in gravitation \citep{Raychaudhuri:1953yv, Kar:2006ms, Vaidya:1999zz, Vaidya:1999zza, Vishveshwara:1970cc, Vishveshwara:1970zz, nrsen, Narlikar:1980gja}. The work of Raychaudhuri opened up the development of BH physics, including singularity theorems. Dynamical BH solutions can be attributed to P.C. Vaidya, while the study involving the stability of BHs was pioneered by C.V. Vishveshwara. In astrophysics and cosmology also, Indian science community had contributed significantly, e.g., in the context of the magnetic Penrose process, quasars, and steady-state cosmology, among others \citep{penroseprocess, quasars, steadystate}. 

Indian scientists have made seminal contributions to developing GW signal detection techniques ~\citep{Sathyaprakash:1991mt} and to constructing accurate theoretical models of expected signals from inspiralling compact binaries using high-order post-Newtonian calculations~\citep{BDIWW95}. This body of work has played a foundational role in enabling the recent detection of GWs and the beginning of the field of GW astronomy. 
Early contributions to the study of gravitational lensing include the development of modeling tools, analysis of lensing by single and multiple lenses, and developing a theoretical framework to combine strong lensing and microlensing images. 

Subsequently, the Indian community also embarked upon the study of the quantum nature of gravity, and there have been works on formal developments of quantum field theory, quantum cosmology, string theory, as well as on loop quantum gravity. Besides, there has been steady growth in the study of astrophysics, cosmology and inflationary scenario. 

 The last few decades observed developments in both theoretical and observational avenues and the Indian science community expanded its horizons as well. This is when Indian scientists started joining and contributing significantly to big projects, e.g., the LIGO Collaboration, the Planck Mission and International Pulsar Timing Array Consortium. There have also been developments in constructing indigenous Indian observatories, e.g., the Astrosat. The detection of GWs by the LIGO detectors, CMB being mapped to unprecedented details by the Planck satellite and multi-wavelength of observations by Astrosat, etc., contributed to the development of a thriving new gravity community in India.

Currently, the Indian science community is actively involved in theoretical research related to various aspects of GW astronomy. In particular, there have been interesting work on tests of GR and the nature of BHs through GWs. Extracting observables from alternative gravity theories and implications for the quantum nature of BHs is another active area of research pursued by several Indian scientists. Indian scientists have also contributed to the construction of shadows by ultra-compact objects in GR and also in modified theories of gravity. Various mathematical properties of the photon sphere, necessary for the construction of BH shadow have also been explored. Application of alternative theories of gravity in the cosmological sector, in particular, the explanation behind the late time acceleration is something Indian researchers are significantly contributing. Indian contributions to the fields of early universe cosmology, inflation and bounce, and quantum cosmology have been persistent over the years, including at current times. Confronting both early and late-time cosmology with the CMB and supernovae measurements is another active area of research.

Presently, the Indian community has an active presence in the international LIGO-Virgo-KAGRA Collaboration. Notable contributions of Indian scientists include the development of accurate theoretical models of expected signals from coalescing compact binaries, improved signal detection algorithms for poorly modeled transient GW signals, search methods for anisotropic stochastic GW backgrounds, tests of GR using GW observations, developing theoretical models of NS equation of state, constraints of the nuclear equation of state, search for continuous waves from spinning NSs in binaries, searches for gravitational lensing of GWs, multimessenger follow-ups for GW events, astrophysics of gamma-ray bursts, etc. Indian groups also have contributed to aspects of enhancing the science capabilities of the upcoming Laser Interferometer Space Antenna (LISA) and in a reformulation of the time-delay interferometry. 

The ongoing LIGO-India project aims to build a third LIGO observatory on Indian soil. Apart from providing a significant enhancement of the science capabilities of the international GW detector network (such as the sky localization of GW events, extraction of polarization, etc.), this challenging project will bring in expertise in several aspects of precision metrology and instrumentation to the Indian community. Its scientific impact extends far beyond GW astrophysics to fundamental physics, particle physics and nuclear physics.
Further, the presence of LIGO-India in the next-generation ground-based GW network will play an important role for multi-messenger astronomy.

The InPTA Consortium is pursuing efforts to establish persistent multi-messenger nano-Hz GW astronomy using the uGMRT. The InPTA experiment has been operational with the upgraded GMRT since  2018, and has become part of IPTA since 2021.  The InPTA collaboration has been monitoring about two dozen pulsars for precision timing and dispersion measurement estimates. InPTA had its first data release in 2022 consisting of pulse arrival times and dispersion measures of 14 millisecond pulsars~\citep{Tarafdar}. The one part in 5 precision of dispersion measure in this data release is unprecedented in any constituent PTA experiment of IPTA so far and promises to significantly enhance the signal-to-noise ratio of detected GW in the upcoming combination of IPTA Data release 3. The low-frequency sensitivity of the uGMRT telescope provides a unique niche in the search for nHz GWs. The InPTA data is being combined with data from other established PTAs for the first detection of nHz GWs. The InPTA observation strategies and experience will also be relevant for the upcoming square kilometer array (SKA) era~\citep{BCJ22}.

\section{Key science questions and goals from an Indian perspective}

\subsection{kHz GW observations using ground-based interferometric detectors:}
Indian community’s continued active participation in the LIGO-Virgo-KAGRA collaboration will ensure its ability to contribute to the big discoveries awaited. LIGO-India detector will play a very important role in the localization of GW sources and thereby play a pioneering role in multimessenger astronomy. The proposed Indian high-energy space telescope Daksha will be an excellent addition to this frontier, complementing GW astronomy and enhancing the prospects of multimessenger astronomy. These efforts should naturally extend to the science of next-generation GW detectors such as Voyager, Einstein Telescope and Cosmic Explorer. 

\subsection{Nano-Hz GW observations using PTAs:}  
The InPTA consortium will continue to establish efforts towards multi-messenger nHz GW astronomy during the SKA era.  InPTA data is now being combined with data from other PTAs, which will help enhance the statistical significance of the recent evidence of nHz GWs.  An ongoing focus is on modeling the expected PTA timing residuals from SMBH binaries in relativistic eccentric orbits and developing approaches to extract observational implications of SMBH binaries in high-frequency radio observations. These efforts are influenced 
by the possibility that a unique blazar OJ 287 may be hosting a SMBH binary that inspirals due to the emission of nHz GWs. To strengthen these efforts, it will be important to develop close collaboration between researchers working on AGNs, radio, and neutrino astronomy and PTAs.

\subsection{Milli-Hz GW observations using space-based interferometric detectors:}   
While continuing its active role in the kHz and nHz GW science, the Indian community should enhance its involvement in the mHz science with LISA, which will be taking data in the next decade. Though there is a small community that contributes to LISA science, this can easily be enhanced. LISA data analysis presents new challenges, and the expertise of the Indian GW community can play an important role in this area. Further, early involvement of the Indian astrophysics community working on SMBHs, galaxy evolution, AGNs, etc, in LISA science is highly desirable. 

\subsection{Deci-Hz GW observations using space or Lunar detectors:}
In the next decade, GW detectors will be observing in the very low frequency (nHz –  mHz) and high-frequency (Hz – kHz) bands. The missing deci-Hz band that lies between LISA and ground-based detectors is of significant astrophysical importance. Deci-Hz GW observations can provide early warning of binary mergers observable by ground-based detectors as well as detect new sources such as intermediate-mass BHs and white dwarf mergers. The international GW community is actively looking at possible concepts (both space and moon-based) for deci-Hz detectors. There is an opportunity for the Indian scientific community to make an early start and to be a leading player in this field. This is also closely aligned with the interest of the Indian Space Program. A detailed study of the feasibility and the science case will be a good starting point.

\subsection{VLBI imaging of BH shadows:}
Another important area where India’s gravity community needs to invest more is the strong-field tests of gravity using present and next-generation EHT-like imaging instruments. Besides data analysis, EHT-related science crucially needs theoretical development, especially that of models that invoke BH mimickers and BHs in alternative theories. A healthy interface with the theoretical gravity community here should help the Indian community in leading this area of research internationally. Despite the challenges involved, it might be also worth for the Indian community to get involved in the millimeter/sub-millimeter radio astronomy efforts. 

\subsection{Theoretical modelling of gravitational phenomena:}
Analytical and numerical modeling of gravitational phenomena is crucial for addressing the open issues in the field. India has traditionally had a strong presence in analytical gravity. Recently, Indian groups have also been involved in numerical simulations, thanks to a new pool of young researchers. Upcoming ground- and space-based GW detectors will require the expected GW signals to be modeled with very high accuracy. This requires a combination of analytical and numerical techniques in GR, including post-Newtonian methods, self-force calculations, BH perturbation theory and numerical relativity. Modeling of some of the more complex astrophysical sources such as NS mergers and core-collapse supernovae or SMBH binaries in gas-rich environments requires general relativistic magnetohydrodynamics in conjunction with sophisticated radiation transfer, nuclear physics models, etc. Accurate modeling of shadows of BHs and ultracompact objects for the next-generation VLBI experiments also poses similar challenges and opportunities. A key component of the theoretical modeling for interpreting VLBI images and movies will also require developing expertise in general relativistic magnetohydrodynamics and general relativistic radiative transfer codes in the Indian community.

\subsection{Novel data analysis and modelling techniques:}
In the above-mentioned areas, development of novel algorithms for theoretical modelling and/or data analysis is extremely important. As Artificial Intelligence and Machine Learning is going to revolutionize the the computing paradigm, it is important that the Indian community is up-to-date about these novel techniques and takes part in pushing these frontiers further. A detailed discussion on this can be found in the {\it Computational Astrophysics} chapter of the Vision Document~\citep{CompAstroChap}.

\section{ Recommendations and priorities}
\begin{enumerate}
\item LIGO-India is a golden opportunity for the Indian physics and astronomy community to be a major player in a rapidly growing frontier. \emph{Dedicated efforts should be made towards the construction, commissioning and timely operation of LIGO-India to maximize its impact on the worldwide network of GW detectors.} At the same time, there should be a special research funding program to create a sizeable user community in India to extract science from the upcoming LIGO-India data and to enable the contribution of the wider Indian scientific community to the development of high-end technologies for LIGO-India upgrades and next-generation GW detectors. 
\item \emph{Adequate uGMRT observing time should be provided to ensure continued critical InPTA contributions to the IPTA consortium.} Further, upgrading of current facilities to fulfill PTA science potential during the era of the SKA should be encouraged. This can, for example, include an Asian VLBI effort that employs uGMRT as an anchoring station.
\item In the next decade, LISA will inaugurate mHz GW observations, offering a new window for astrophysics. \emph{Indian GW/astrophysics communities should step up their involvement in the LISA data analysis and astrophysics.}
\item Some of the proposed space-based detector concepts for deci-Hz GW astronomy are closely aligned with the interests of the Indian Space Program. Since these efforts are in the nascent stages, \emph{there is an opportunity for the Indian community to be a major player in deci-Hz GW astronomy.}

\item High-performance/high-throughput computing is an essential part of modern astronomy. Computational challenges involve the analysis of large volumes of data and the computation of complex theoretical models to interpret the observations. \emph{Significant computing resources will have to be generated within India for playing leadership roles in these activities. A transparent and user-friendly approach towards creating and sharing computing resources within the country is badly needed}.

\item Theoretical modeling of gravitational phenomena using analytical and numerical tools will play a central role in GW observations, VLBI imaging, etc. \emph{Indian community's historical strength in theoretical gravity research should be sustained and nurtured.} There should be a dedicated funding mechanism for organizing regular pedagogical schools for attracting and training students at different levels (similar to the SERC schools in Theoretical High Energy Physics). These should be supplemented with workshops, symposia and networking meetings for researchers.  

\item \emph{An attractive and internationally competitive postdoctoral program is crucial for establishing a dynamic research community, and for arresting the brain drain} (similar to, e.g., NASA Einstein/Hubble fellowships).

\section{Outlook}

Gravitational physics has become a highly interdisciplinary field that connects with several areas of physics such as high energy physics, nuclear physics and cosmology, and to almost all branches of astronomy. This interplay means that developments in these areas directly influence research in gravitational physics and the results of this field significantly impact all these branches of physics. In addition, frontier astronomy projects often develop state-of-the-art technology that can be directly borrowed by the industry. As this century marks the era of multi-messenger astronomy, most of the astronomy mega-projects address research problems in a broad spectrum of astrophysics and gravitational physics. Further, there are huge technology spin-offs from these projects and large-scale training of highly skilled human resources which can be directly borrowed in the industry. Therefore, a close connection with society in terms of the exchange of knowledge and skills is highly desirable for the benefit of society at large. 

The community must actively seek to establish connections with students from undergraduate colleges and schools, through workshops and training programs. By engaging with these young, enthusiastic individuals early in their academic careers, the community can foster a deeper interest in gravitational physics, provide mentorship and guidance, and create pathways for these students to pursue advanced studies and research in the field. The broad appeal of the topics may also be utilized to generate funds, say from private donors, to support building the community, especially supporting young researchers in the early phase of their careers.

One cannot stress enough the importance of diversity and inclusion in science. Groups containing researchers from diverse backgrounds (social, gender, ethnicity, economic class, etc) have the potential to make scientific breakthroughs and in a country like ours, known for diversity, it is only natural to expect that these key aspects are part of every research group.

\end{enumerate}
%\section{Appendix section}
\bibliographystyle{apj}
\bibliography{references}

\begin{thebibliography}{}
\expandafter\ifx\csname natexlab\endcsname\relax\def\natexlab#1{#1}\fi

\bibitem[{Abbott {$et~al$.}(2016{\natexlab{a}})}]{LIGOScientific:2016aoc}
Abbott, B.~P., {$et~al$.} 2016{\natexlab{a}}, Phys. Rev. Lett., 116, 061102

\bibitem[{Abbott {$et~al$.}(2016{\natexlab{b}})}]{GW150914TGR}
---. 2016{\natexlab{b}}, Phys. Rev. Lett., 116, 221101, [Erratum:
  Phys.Rev.Lett. 121, 129902 (2018)]

\bibitem[{Abbott {$et~al$.}(2017{\natexlab{a}})}]{GW170817}
---. 2017{\natexlab{a}}, Phys. Rev. Lett., 119, 161101

\bibitem[{Abbott {$et~al$.}(2017{\natexlab{b}})}]{MMA}
---. 2017{\natexlab{b}}, Astrophys. J. Lett., 848, L12

\bibitem[{Abbott {$et~al$.}(2018)}]{LVKLRR}
---. 2018, Living Rev. Rel., 21, 3

\bibitem[{Agathos {$et~al$.}(2015)Agathos, Meidam, Del~Pozzo, Li, Tompitak,
  Veitch, Vitale, \& Van Den~Broeck}]{Agathos:2015uaa}
Agathos, M., Meidam, J., Del~Pozzo, W., {$et~al$.} 2015, Phys. Rev. D, 92,
  023012

\bibitem[{Agazie {$et~al$.}(2023)}]{NANOGrav:2023det}
Agazie, G., {$et~al$.} 2023, Astrophys. J. Lett., 951, L8

\bibitem[{Antoniadis {$et~al$.}(2023)}]{InPTAEPTA}
Antoniadis, J., {$et~al$.} 2023, Astron. Astrophys., 678, A50

\bibitem[{{Ayzenberg} {$et~al$.}(2023){Ayzenberg}, {Blackburn}, {Brito},
  {Britzen}, {Broderick}, {Carballo-Rubio}, {Cardoso}, {Chael}, {Chatterjee},
  {Chen}, {Cunha}, {Davoudiasl}, {Denton}, {Doeleman}, {Eichhorn}, {Eubanks},
  {Fang}, {Foschi}, {Fromm}, {Galison}, {Ghosh}, {Gold}, {Gurvits}, {Hadar},
  {Held}, {Houston}, {Hu}, {Johnson}, {Kocherlakota}, {Natarajan}, {Olivares},
  {Palumbo}, {Pesce}, {Rajendran}, {Roy}, {Saurabh}, {Shao}, {Tahura}, {Tamar},
  {Tiede}, {Vincent}, {Visinelli}, {Wang}, {Wielgus}, {Xue}, {Yakut}, {Yang},
  \& {Younsi}}]{Ayzenberg+2023}
{Ayzenberg}, D., {Blackburn}, L., {Brito}, R., {$et~al$.} 2023, arXiv e-prints,
  arXiv:2312.02130

\bibitem[{Blanchet {$et~al$.}(1995)Blanchet, Damour, Iyer, Will, \&
  Wiseman}]{BDIWW95}
Blanchet, L., Damour, T., Iyer, B.~R., Will, C.~M., \& Wiseman, A.~G. 1995,
  Phys. Rev. Lett., 74, 3515

\bibitem[{{Blandford} \& {Znajek}(1977)}]{Blandford+1977}
{Blandford}, R.~D., \& {Znajek}, R.~L. 1977, \mnras, 179, 433

\bibitem[{Coulter {$et~al$.}(2017)}]{SwopeEMCounterpart}
Coulter, D.~A., {$et~al$.} 2017, Science, 358, 1556

\bibitem[{{Doeleman} {$et~al$.}(2023){Doeleman}, {Barrett}, {Blackburn},
  {Bouman}, {Broderick}, {Chaves}, {Fish}, {Fitzpatrick}, {Freeman}, {Fuentes},
  {G{\'o}mez}, {Haworth}, {Houston}, {Issaoun}, {Johnson}, {Kettenis},
  {Loinard}, {Nagar}, {Narayanan}, {Oppenheimer}, {Palumbo}, {Patel}, {Pesce},
  {Raymond}, {Roelofs}, {Srinivasan}, {Tiede}, {Weintroub}, \&
  {Wielgus}}]{Doeleman+2023}
{Doeleman}, S.~S., {Barrett}, J., {Blackburn}, L., {$et~al$.} 2023, Galaxies,
  11, 107

\bibitem[{{Event Horizon Telescope Collaboration}
  {$et~al$.}(2019{\natexlab{a}})}]{EHTC+2019a}
{Event Horizon Telescope Collaboration}, {$et~al$.} 2019{\natexlab{a}},
  Astrophys. J. Lett., 875, L1

\bibitem[{{Event Horizon Telescope Collaboration}
  {$et~al$.}(2019{\natexlab{b}})}]{EHTC+2019f}
---. 2019{\natexlab{b}}, Astrophys. J. Lett., 875, L6

\bibitem[{{Event Horizon Telescope Collaboration}
  {$et~al$.}(2022{\natexlab{a}})}]{EHTC+2022a}
---. 2022{\natexlab{a}}, Astrophys. J. Lett., 930, L12

\bibitem[{{Event Horizon Telescope Collaboration}
  {$et~al$.}(2022{\natexlab{b}})}]{EHTC+2022f}
---. 2022{\natexlab{b}}, Astrophys. J. Lett., 930, L17

\bibitem[{Foucart(2020)}]{Foucart:2020ats}
Foucart, F. 2020, Front. Astron. Space Sci., 7, 46

\bibitem[{Ghez {$et~al$.}(2008)}]{Ghez:2008ms}
Ghez, A.~M., {$et~al$.} 2008, Astrophys. J., 689, 1044

\bibitem[{Gillessen {$et~al$.}(2009)Gillessen, Eisenhauer, Trippe, Alexander,
  Genzel, Martins, \& Ott}]{Gillessen:2008qv}
Gillessen, S., Eisenhauer, F., Trippe, S., {$et~al$.} 2009, Astrophys. J., 692,
  1075

\bibitem[{Gupta {$et~al$.}(2017)}]{uGMRT}
Gupta, Y., {$et~al$.} 2017, Current Science, 113, 707

\bibitem[{{Hoyle} {$et~al$.}(1993){Hoyle}, {Burbidge}, \&
  {Narlikar}}]{steadystate}
{Hoyle}, F., {Burbidge}, G., \& {Narlikar}, J.~V. 1993, \apj, 410, 437

\bibitem[{Iyer {$et~al$.}(2011)Iyer, Souradeep, Unnikrishnan, Dhurandhar, Raja,
  \& Sengupta}]{LIGOIndiaProposal:2011}
Iyer, B., Souradeep, T., Unnikrishnan, C., {$et~al$.} 2011, {LIGO} Technical
  Document LIGO-M1100296-v2

\bibitem[{{Johnson} {$et~al$.}(2020){Johnson}, {Lupsasca}, {Strominger},
  {Wong}, {Hadar}, {Kapec}, {Narayan}, {Chael}, {Gammie}, {Galison}, {Palumbo},
  {Doeleman}, {Blackburn}, {Wielgus}, {Pesce}, {Farah}, \&
  {Moran}}]{Johnson+2020}
{Johnson}, M.~D., {Lupsasca}, A., {Strominger}, A., {$et~al$.} 2020, Science
  Advances, 6, eaaz1310

\bibitem[{{Johnson} {$et~al$.}(2023){Johnson}, {Akiyama}, {Blackburn},
  {Bouman}, {Broderick}, {Cardoso}, {Fender}, {Fromm}, {Galison}, {G{\'o}mez},
  {Haggard}, {Lister}, {Lobanov}, {Markoff}, {Narayan}, {Natarajan}, {Nichols},
  {Pesce}, {Younsi}, {Chael}, {Chatterjee}, {Chaves}, {Doboszewski}, {Dodson},
  {Doeleman}, {Elder}, {Fitzpatrick}, {Haworth}, {Houston}, {Issaoun},
  {Kovalev}, {Levis}, {Lico}, {Marcoci}, {Martens}, {Nagar}, {Oppenheimer},
  {Palumbo}, {Ricarte}, {Rioja}, {Roelofs}, {Thresher}, {Tiede}, {Weintroub},
  \& {Wielgus}}]{Johnson+2023}
{Johnson}, M.~D., {Akiyama}, K., {Blackburn}, L., {$et~al$.} 2023, Galaxies,
  11, 61

\bibitem[{Joshi {$et~al$.}(2022)}]{BCJ22}
Joshi, B., {$et~al$.} 2022, J. Astrophys. Astron., 43, 98

\bibitem[{Kar \& SenGupta(2007)}]{Kar:2006ms}
Kar, S., \& SenGupta, S. 2007, Pramana, 69, 49

\bibitem[{{Kembhavi} \& {Narlikar}(1999)}]{quasars}
{Kembhavi}, A.~K., \& {Narlikar}, J.~V. 1999, {Quasars and active galactic
  nuclei : an introduction}

\bibitem[{{Kocherlakota} {$et~al$.}(2021)}]{Kocherlakota+2021}
{Kocherlakota}, P., {$et~al$.} 2021, Phys. Rev. D, 103, 104047

\bibitem[{{Lasota} {$et~al$.}(2014){Lasota}, {Gourgoulhon}, {Abramowicz},
  {Tchekhovskoy}, \& {Narayan}}]{Lasota+2014}
{Lasota}, J.~P., {Gourgoulhon}, E., {Abramowicz}, M., {Tchekhovskoy}, A., \&
  {Narayan}, R. 2014, \prd, 89, 024041

\bibitem[{Lattimer \& Prakash(2001)}]{Lattimer:2000nx}
Lattimer, J.~M., \& Prakash, M. 2001, Astrophys. J., 550, 426

\bibitem[{Margalit(2019)}]{Margalit:2019ssg}
Margalit, B. 2019, Annals Phys., 410, 167925

\bibitem[{Narlikar(1980)}]{Narlikar:1980gja}
Narlikar, V.~V. 1980, in {Einstein Centenary Symposium}, 3--17

\bibitem[{{Nobel Press Release}(2018)}]{GWnobel}
{Nobel Press Release}. 2018, {{\rm Gravitational waves finally captured}},
  \url{https://www.nobelprize.org/prizes/physics/2017/press-release/}

\bibitem[{{Nobel Press Release}(2019)}]{Cosmonobel}
---. 2019, {{\rm New perspectives on our place in the universe }},
  \url{https://www.nobelprize.org/prizes/physics/2019/press-release/}

\bibitem[{{Nobel Press Release}(2020)}]{BHnobel}
---. 2020, {{\rm Black holes and the Milky Way’s darkest secret}},
  \url{https://www.nobelprize.org/prizes/physics/2020/press-release//}

\bibitem[{{Psaltis} {$et~al$.}(2020)}]{Psaltis+2020}
{Psaltis}, D., {$et~al$.} 2020, Phys. Rev. Lett., 125, 141104

\bibitem[{Punturo {$et~al$.}(2010)}]{ET}
Punturo, M., {$et~al$.} 2010, Class. Quant. Grav., 27, 194002

\bibitem[{Raychaudhuri(1955)}]{Raychaudhuri:1953yv}
Raychaudhuri, A. 1955, Phys. Rev., 98, 1123

\bibitem[{Reitze {$et~al$.}(2019)}]{CE:2019iox}
Reitze, D., {$et~al$.} 2019, Bull. Am. Astron. Soc., 51, 035

\bibitem[{{Ripperda} {$et~al$.}(2022){Ripperda}, {Liska}, {Chatterjee},
  {Musoke}, {Philippov}, {Markoff}, {Tchekhovskoy}, \&
  {Younsi}}]{Ripperda+2022}
{Ripperda}, B., {Liska}, M., {Chatterjee}, K., {$et~al$.} 2022, \apjl, 924, L32

\bibitem[{Saleem {$et~al$.}(2022)}]{Saleem:2021iwi}
Saleem, M., {$et~al$.} 2022, Class. Quant. Grav., 39, 025004

\bibitem[{Sathyaprakash \& Dhurandhar(1991)}]{Sathyaprakash:1991mt}
Sathyaprakash, B.~S., \& Dhurandhar, S.~V. 1991, Phys. Rev. D, 44, 3819

\bibitem[{Sen \& Eddington(1934)}]{nrsen}
Sen, N.~R., \& Eddington, A. 1934, Monthly Notices of the Royal Astronomical
  Society, 94, 550

\bibitem[{Sharma {$et~al$.}(2025)Sharma, Vaidya, Wadadekar, Bagla, Chatterjee,
  Hanasoge, Kumar, Mukherjee, Philip, \& Singh}]{CompAstroChap}
Sharma, P., Vaidya, B., Wadadekar, Y., {$et~al$.} 2025, Computational
  Astrophysics, Data Science \& AI/ML in Astronomy: A Perspective from Indian
  Community, arXiv:2501.03876

\bibitem[{Steiner {$et~al$.}(2013)Steiner, Lattimer, \& Brown}]{Steiner:2012xt}
Steiner, A.~W., Lattimer, J.~M., \& Brown, E.~F. 2013, Astrophys. J. Lett.,
  765, L5

\bibitem[{Tarafdar {$et~al$.}(2022)}]{Tarafdar}
Tarafdar, P., {$et~al$.} 2022, Publ. Astron. Soc. Austral., 39, e053

\bibitem[{{Vagnozzi} {$et~al$.}(2023){Vagnozzi}, {Roy}, {Tsai}, {Visinelli},
  {Afrin}, {Allahyari}, {Bambhaniya}, {Dey}, {Ghosh}, {Joshi}, {Jusufi},
  {Khodadi}, {Walia}, {{\"O}vg{\"u}n}, \& {Bambi}}]{Vagnozzi+2023}
{Vagnozzi}, S., {Roy}, R., {Tsai}, Y.-D., {$et~al$.} 2023, Class. Quantum
  Grav., 40, 165007

\bibitem[{Vaidya(1999{\natexlab{a}})}]{Vaidya:1999zz}
Vaidya, P.~C. 1999{\natexlab{a}}, Gen. Rel. Grav., 31, 119

\bibitem[{Vaidya(1999{\natexlab{b}})}]{Vaidya:1999zza}
---. 1999{\natexlab{b}}, Gen. Rel. Grav., 31, 121

\bibitem[{Vishveshwara(1970{\natexlab{a}})}]{Vishveshwara:1970zz}
Vishveshwara, C.~V. 1970{\natexlab{a}}, Nature, 227, 936

\bibitem[{Vishveshwara(1970{\natexlab{b}})}]{Vishveshwara:1970cc}
---. 1970{\natexlab{b}}, Phys. Rev. D, 1, 2870

\bibitem[{{Wagh} {$et~al$.}(1985){Wagh}, {Dhurandhar}, \&
  {Dadhich}}]{penroseprocess}
{Wagh}, S.~M., {Dhurandhar}, S.~V., \& {Dadhich}, N. 1985, \apj, 290, 12

\end{thebibliography}
\end{document}